\documentclass[
  prd,
  reprint,
  floats,
  floatfix,
  amssymb,
  superscriptaddress,
  nofootinbib
]{revtex4-2}

\usepackage{graphicx}
\usepackage{dcolumn}
\usepackage{bm}
\usepackage{amsmath}
\usepackage[colorlinks=true,linkcolor=blue,citecolor=blue,urlcolor=blue]{hyperref}
\usepackage{dsfont}
\usepackage{subcaption}
\usepackage{xcolor}
\usepackage{ulem}

\newcommand\comment[1]{}

\begin{document}

\title{Fast Fourier Transform evaluation of the Fresnel integral for gravitational-wave lensing}

\author{Nino Ephremidze}
\email{nino\_ephremidze@g.harvard.edu}
\affiliation{Department of Physics, Harvard University, Cambridge, MA 02138, USA}

\author{Marc Kamionkowski}
\email{kamion@jhu.edu}
\affiliation{William H.\ Miller III Department of Physics \& Astronomy, Johns Hopkins University, 3400 N.\ Charles St., Baltimore, MD 21218, USA}

\author{Cora Dvorkin}
\email{cdvorkin@g.harvard.edu}
\affiliation{Department of Physics, Harvard University, Cambridge, MA 02138, USA}

\setcounter{footnote}{0}
\def\thefootnote{\arabic{footnote}}

\begin{abstract}
Gravitational waves (GWs) exhibit wave-optics effects when their wavelength is comparable to the scale of the gravitational lens. This may occur in lensing from galactic subhalos in GWs emitted by binary black-hole mergers, and is gaining interest as a novel probe of dark matter. Predictions for observables in these cases ultimately rely on evaluating a Fresnel integral that quantifies the effect of lensing on the amplitude of a GW at a given frequency. However, numerical evaluation of this Fresnel integral is tricky, and several algorithms and publicly available codes that implement it have been developed. Here, we show that the dependence of this integral on the lens position can be written as a two-dimensional Fourier transform. Modern FFT techniques then enable rapid evaluation at all-sky positions simultaneously for general lenses without symmetry. Vectorization of FFT routines allows for derivatives with respect to model parameters to be obtained with only incremental additional computational cost. If the lens is axisymmetric, further speedups can be achieved with recently developed techniques for non-uniform fast Hankel transforms. To demonstrate, we make available \textit{Fresnel Integral Optimization with Non-uniform trAnsforms} (\texttt{FIONA}), an efficient and accurate code that is significantly faster than current methods for dense source grids, reaching 2--3 orders of magnitude speedups for $\sim 10^6$ GW-emitting points. As part of \texttt{FIONA}, we developed code that provides vectorized non-uniform fast Hankel transforms that may have other uses (e.g., calculation of cosmological two-point correlation functions) beyond those considered here.
\end{abstract}

\maketitle

\section{Introduction}

Gravitational waves can, like electromagnetic waves, be gravitationally lensed \cite{Ohanian:1974ys,Wang:1996as,Nakamura:1997sw,Sereno:2010dr,Meena:2019ate,Mukherjee:2019wcg,Ezquiaga:2020spg,Fairbairn:2022xln}.  In almost all astrophysical cases, lensing of electromagnetic waves occurs in the geometrical-optics limit: the wavelengths are tiny compared with the lens, and so light rays propagate along the smallest time-delay trajectory. However, it is much more likely that for gravitational waves, where the emitters and lenses may have comparable masses, wave-optic effects may be important \cite{Deguchi:1986zz,DePaolis:2002tw,Takahashi:2003ix}. For example, GWs from the merger of $\sim10^7\,M_\odot$ black holes may conceivably be lensed by galactic subhalos of comparable mass \cite{Oguri:2020ldf,Gao:2021sxw,Guo:2022dre,Caliskan:2023zqm,Cheung:2024ugg}, and stellar-mass lenses may be lensed by stars or other stellar-remnant black holes \cite{Cremonese:2021puh}.

The effects of lensing of a GW source are quantified in terms of a Fresnel integral (detailed below) that multiplies the Fourier-space waveform.  This Fresnel integral $F(w,{\bf y})$ is written as an integral over the entire lens plane, and it depends on the (scaled) GW frequency $w$ and dimensionless position ${\bf y}$ of the source---i.e., three independent variables most generally, or two if the lens is axisymmetric.  The integrand oscillates very rapidly, making direct evaluation tricky.  There are analytic solutions for a handful of axisymmetric lenses.  However, even these ``analytic'' solutions are in terms of obscure hypergeometric functions and/or infinite sums, and evaluating these numerically is not trivial. The Fresnel integral can be evaluated at high frequencies with a stationary-phase approximation and approximated with Taylor expansions at low frequencies.

A variety of numerical approaches have been developed to deal with intermediate frequencies.  One \cite{Ulmer:1994ij,Tambalo:2022plm,Villarrubia-Rojo:2024xcj} starts in the time domain, where the transfer function can be written as an integral over contours of equal time delay in the lens plane.  The results are then Fourier-transformed back to frequency space to obtain $F(w,{\bf y})$.  This contour approach \cite{Villarrubia-Rojo:2024xcj} efficiently computes the Fresnel integral but only at a single point in the lens plane.  The Picard-Lefschetz approach \cite{Feldbrugge:2019fjs,Feldbrugge:2020ycp} identifies contours in the complex plane where rapidly-oscillating integrands become exponentially decaying. This works well at high frequencies but becomes less efficient at lower frequencies, where the contours must extend farther into the complex plane.  Another option \cite{Grillo:2018qjt} notes that the integral over the lens plane can be written as a convolution, which then amounts to multiplication in the Fourier domain. FFTs then allow simultaneous computation of results over the entire lens plane for each frequency. Implementing the forward and reverse transforms, however, is numerically challenging.  Finally, Ref.~\cite{Caliskan:2023zqm} developed a rapid way to precisely interpolate numerical results by interpolating the smoothly varying amplitude and phase of $F(w,{\bf y})$ on $w$.

Although detection of lensed gravitational waves may be some way off in the future, the Fresnel integral appears in a variety of other areas of science and engineering, including femtolensing \cite{Ulmer:1994ij}, astrophysical radio waves as they propagate through the interstellar medium \cite{Grillo:2018qjt}, starshade design \cite{Barnett2021}, and imaging technologies \cite{Yan:2025}.  Efforts to think anew about this integral are thus well motivated, not just for computational speed and efficiency, but also for the insight they may provide into existing schemes.

Here, we adapt a method \cite{Barnett2021} to calculate diffraction patterns for star shades (and also in 3d imaging \cite{Li2020FresnelZonePlate})
to the case of GW lensing.  The spatial dependence of the Fresnel integral can be expressed as a two-dimensional Fourier transform, and we then leverage powerful fast Fourier transform (FFT) techniques.  In particular, a non-uniform FFT (NUFFT) \cite{BarnettMaglandKlinteberg2019,Barnett2021a} allows for rapid evaluation of the Fourier integral over a finite interval, mitigating the notorious problems with ringing that arise in approximating a Fourier integral by a discrete FFT.  We then window the potential so that problems associated with integration at large distances, where the integrand varies most rapidly, are avoided.  This ``brute-force'' approach is conceptually simple and straightforward to code, and yields results for the entire lens plane in one fell swoop.  Moreover, the vectorization capabilities of the NUFFT allow derivatives with respect to the lens-plane position and/or model parameters to be obtained with only an incremental increase in computational time.  If the lens is axisymmetric, the two-dimensional NUFFT can be replaced by a one-dimensional non-uniform fast Hankel transform (NUFHT) \cite{beckman2024}, which then provides results in the entire $w$-$y$ plane even more quickly.

In the following, we present the basic ideas behind the calculation and describe some details of the codes we have written to test this approach.  Our aim with these codes is to show there are no show-stoppers in the numerical implementation of the ideas.  We provide our codes (in Python and C) for those interested to use, explore, and/or develop further\footnote{\url{https://github.com/ninoephremidze/FIONA.git} \newline \url{https://github.com/marckamion/FHTFresnel} \newline \url{https://github.com/marckamion/fresnel2d} \newline \url{https://github.com/marckamion/NUFHT}}.  Although these codes contain some elementary tricks to accelerate the calculation and are reasonably fast, further optimization is certainly possible with additional effort.

The organization of this paper is as follows:  In Section \ref{sec:GWlensing} we review the theory of wave-optics effects in gravitational lensing and the Fresnel diffraction integral that is the subject of this paper.  Section \ref{sec:FTapproach} presents the approach to evaluating the Fresnel diffraction integral in terms of non-uniform Fourier or Hankel transforms, firstly for general two-dimensional lens geometries without symmetry, and then also for axisymmetric lenses.  This Section also describes how the algorithms are implemented numerically.  Section \ref{sec:code_perf} evaluates the performance of the code we have written, comparing it with {\tt GLoW} \cite{Villarrubia-Rojo:2024xcj}.  Section \ref{sec:results} shows results of the evaluation of the Fresnel diffraction integral for a variety of lens configurations, and Section \ref{sec:conclusions} provides concluding remarks.  One Appendix demonstrates the numerical agreement between our results and those of {\tt GLoW}, and another provides details about the lens models used in this work.

\section{Gravitational wave lensing}
\label{sec:GWlensing}

In this section, we briefly review the formalism of gravitational lensing.

The lensed waveform in the frequency domain \( \tilde{h}_{\rm obs}(f) \) is related to the unlensed waveform \( \tilde{h}_{\rm src}(f) \) by
\begin{equation}
    \tilde{h}_{\rm obs}(f) = F(w,{\bf y})\,\tilde{h}_{\rm src}(f),
\end{equation}
where \(F(w,{\bf y})\) is the complex amplification factor and \({\bf y}\) is the (dimensionless) source position on the sky scaled by a characteristic angular scale of the lens. The dimensionless frequency parameter \(w\) is defined as
\begin{equation}
    w \equiv 4\pi\,f\,\frac{GM(1+z_l)}{c^3},
\end{equation}
where \(f\) is the GW frequency, \(M\) is a characteristic lens mass scale, \(z_l\) is the lens redshift, and \(c\) is the speed of light. In this notation, \(w\gg1\) corresponds to the geometric optics regime, while \(w\lesssim1\) captures diffraction effects.

In the thin‐lens approximation, \(F(w,{\bf y})\) is given by the Fresnel–Kirchhoff diffraction integral,
\begin{equation}
    F(w,{\bf y}) = \frac{w}{2\pi i}\int_{\mathbb{R}^2} \mathrm{d}^2x \,
    \exp\!\left\{i\,w\left[\frac{1}{2}\left|\mathbf{x}-\mathbf{y}\right|^2-\psi(\mathbf{x})\right]\right\},
    \label{eq:fresnel_integral}
\end{equation}
where \(\mathbf{x}\) are the dimensionless coordinates on the lens plane normalized to a characteristic angular scale of the lens, and \(\psi(\mathbf{x})\) is the projected dimensionless lensing potential. The potential \(\psi(\mathbf{x})\) is related to the physical surface mass density \(\Sigma(\mathbf{x})\) via the Poisson equation,
\begin{equation}
    \nabla^2_{\bf x}\,\psi(\mathbf{x}) = 2\,\kappa(\mathbf{x}), \quad 
    \kappa(\mathbf{x})\equiv \frac{\Sigma(\mathbf{x})}{\Sigma_{\rm crit}},
\end{equation}
where \(\kappa\) is the convergence and \(\Sigma_{\rm crit}\) is the critical surface density. Multiple image positions in the geometric optics limit correspond to stationary points of the time delay surface,
\begin{equation}
    \tau(\mathbf{x},{\bf y}) \equiv \tfrac{1}{2}\left|\mathbf{x}-\mathbf{y}\right|^2-\psi(\mathbf{x}),
\end{equation}
which obey Fermat’s principle \(\nabla_{\bf x}\tau=0\).

\section{Fourier Transform Approach}
\label{sec:FTapproach}

In this section, we describe the algorithms developed to compute the Fresnel integral involved in the amplification factor $F(w,\mathbf{y})$.

\subsection{General Lens}
\label{sec:general_lens}

The key insight comes by re-writing Eq.~(\ref{eq:fresnel_integral}) as
\begin{eqnarray}
     F(w,{\bf y}) &=& \frac{w}{2 \pi i} e^{i wy^2/2} \int \, d^2x\, e^{i w {\bf x}\cdot{\bf y}} \nonumber \\
     & & \ \ \ \times \exp \left\{ iw \left[x^2/2 - \psi({\bf x}) \right] \right\}.
\label{eqn:2dffta}     
\end{eqnarray}
This shows that the spatial dependence of the integral, for some fixed $w$ is just the 2D Fourier transform of $\exp \left\{ iw \left[x^2/2 - \psi\left({\bf x}\right) \right] \right\}$.  The calculation then reduces to a single two-dimensional numerical Fourier transform, albeit for an extremely rapidly oscillating function.

The extraordinarily rapid variation of the integrand at large distances $|{\bf x}|$ provides the principal difficulty with numerical evaluation. To deal with this, we choose an integration domain $R$ much larger than the characteristic lens scale, so that the lens potential $\psi(\mathbf{x}) \approx 0$ for $|x| > R$ (or, if the lens potential diverges, we window the potential so that $\psi({\bf x})=0$ for $|{\bf x}|>R$). We then use the fact that $F(w,{\bf y})=1$ if $\psi({\bf x})=0$ everywhere to re-write Eq.~(\ref{eqn:2dffta}) as
\begin{eqnarray}
     F(w,{\bf y}) &=& 1 + \frac{w}{2 \pi i}e^{i wy^2/2} \int_{x<R} \, d^2x\, e^{i w {\bf x}\cdot{\bf y}} \nonumber \\    & & \ \ \ \ \times e^{iw x^2/2}
  \left( e^{-iw\psi({\bf x})} -1 \right).
\label{eqn:2dfft}     
\end{eqnarray}
This re-formulation effectively removes the Fresnel oscillations of the free-space ($\psi(\mathbf{x})=0$) integrand coming from the regions outside the quadrature. The oscillations in the remaining integrand are not quite so dramatic and are more amenable to numerical integration.

Although the remaining integral over $x<R$ can be approximated numerically using a traditional FFT, the uniform spacing of the abscissas in this approach leads to notorious ringing artifacts. To mitigate these problems, we instead use a non-uniform FFT (NUFFT) to evaluate the integrals with Gauss-Legendre quadrature.

More precisely, we write the integral in Eq.~(\ref{eqn:2dfft}) as
\begin{equation}
     I(w,{\bf y}_\alpha) = \sum_{\beta} W_\beta e^{iw{\bf x}_\beta \cdot {\bf y_\alpha}}  e^{i wx_\beta^2/2} \left( e^{-i w \psi({\bf x}_\beta)} -1 \right),
\label{eqn:GLquadrature}     
\end{equation}
Here ${\bf x}_\alpha$ are a set of $n_{\rm gl}^2$ points with coordinates $x_1$ taken to be $n_{\rm gl}$ Gauss-Legendre (GL) roots from $-R$ to $R$ and similarly for $x_2$.  The weight $W_\beta$ is then the product of the GL weights for $x_1$ and $x_2$.  The sum in Eq.~(\ref{eqn:GLquadrature}) is evaluated for a two-dimensional grid of points ${\bf y_\alpha}$ uniformly spaced in $((-R,R),(-R,R))$.  This is accomplished for all ${\bf y}_\alpha$ simultaneously with {\tt FINUFFT}'s type-1 NUFFT\footnote{\url{https://github.com/flatironinstitute/finufft.git}}.  The number of GL points must be chosen so that the oscillations in the factors are resolved, which requires the spacing in $x_1$ (and also $x_2$) to be (for $e^{-iwx^2/2}$) $\Delta (wx^2)/2 \lesssim w x\Delta x \simeq 1$ for $x<R$, or $\Delta x\lesssim (wR)^{-1}$.  We thus take $n_{\rm gl}\simeq 2 w R^2$.

Other quadrature choices are possible.  For example, one could write the integral $\int d^2x = \int\, x\, dx \int \, d\theta$ and evaluate the $x$ integral with GL weights and the $\theta$ integral with a trapezoidal rule.  To ensure a reasonably uniform density of points everywhere in the integration region, the number of points in $\theta$ should be proportional to $x$, which adds a slight (although not showstopping) level of complication.  Since the integrand at large $w$ is focused around extrema of $x^2/2-\psi({\bf x})$, it should also be possible, at larger $w$, to put more points near these points.

Before continuing, we comment briefly on the approach of Ref.~\cite{Grillo:2018qjt}.  This relies on the observation that the integral in Eq.~(\ref{eq:fresnel_integral}) is a convolution of $\exp[i w |{\bf x}-{\bf y}|^2/2]$ and $\exp[-iw \psi({\bf x})]$, which is equivalent to multiplication in Fourier space.  The Fourier transform of the first factor is $(2\pi i/w)\exp[-i w k^2/2]$.  The second factor (that depends on the potential $\psi({\bf x})$) can be evaluated numerically using the NUFFT, as above. It is actually a bit easier than the FFT we evaluate for Eq.~(\ref{eqn:2dffta}) since $\psi({\bf x}) \to 0$ at large $|{\bf x}|$.  However, the function to be inverse Fourier transformed has the rapidly oscillating factor $\exp[-i w k^2/2]$, which is thus as difficult as the integral in Eq.~(\ref{eqn:2dffta}).  The computation thus requires two (rather than one) FFTs, one which is a bit easier than the one we do in Eq.~(\ref{eqn:2dffta}) but the other equivalent in difficulty.

\subsection{Axisymmetric Lens}
\label{sec:axisym}

The approach discussed above applies to the most general lens and, of course, also works for axisymmetric lenses.  If, however, the lens is axisymmetric, then the FFTs can be replaced with fast Hankel transforms, allowing for even greater computational efficiency.

If the lens is axisymmetric, then $\psi({\bf x})=\psi(x)$ depends only on the magnitude $x=|{\bf x}|$.  If so, then Eq.~(\ref{eqn:2dfft}) can be written as
\begin{eqnarray}
     F(w, y) &=& \frac{w}{2 \pi i} e^{i wy^2/2} \int_0^\infty \, x dx\, e^{iw \left[x^2/2 - \psi(x) \right]} J_0(w x y) \nonumber \\
     &=& 1 + \frac{w}{2\pi i}e^{iwy^2/2} I(w,y),
\label{eqn:fht}
\end{eqnarray}
with
\begin{equation}
     I(w,y) = \int_0^\infty \, x\,dx\, e^{iwx^2/2} \left( e^{-i \psi(x)}-1 \right)  J_0(w x y).
\label{eqn:I}
\end{equation}
This is the Hankel transform of $\exp \left\{ iw \left[x^2/2 - \psi(x) \right] \right\}$, which maps the function of $x$ to a function of $wy$. Algorithms for fast Hankel transforms optimized for functions that vary smoothly on logarithmic scales \cite{Talman1978,Hamilton:1999uv} have been widely used in cosmology, but the integrand in Eq.~(\ref{eqn:I}) varies rapidly. We therefore use a more recently developed non-uniform FHT (NUFHT) \cite{beckman2024}.  The NUFHT can be used to approximate the integral with a Gauss-Legendre quadrature, and---as with the NUFFT---the non-uniform spacing of the GL roots mitigates ringing in the traditional FHT.

We window the potential so that $\psi(x) \to 0$ at $x>R$, for some large $R$, in which case the integral $I(w,y)$ can be approximated (for some fixed value of $w$) as
\begin{equation}
     I(w,y_j) = \sum_{k=1}^{n_{\rm gl}} c_k J_0\left( x_k w y_j\right),
\label{eqn:Isum}     
\end{equation}
where $c_k = x_k w_k e^{iw x_k^2/2} \left(e^{-i\psi(x_k)} -1 \right)$, and $x_k$ and $w_k$ are the $n_{\rm gl}$ GL roots and weights, respectively. We take GL weights (for integrals from $-1$ to 1) and re-scale them to span the range 0 to $R$.  The NUFHT is evaluated using the algorithm of Ref.~\cite{beckman2024}.


\subsection{Implementation}

In practice, we implement a few steps to make the approach accurate and efficient over a wide range of frequencies. The precision of our computation is controlled by the number of Gauss-Legendre nodes per axis $n_{\text{gl}}$. At large $w$, the integrand becomes increasingly oscillatory, requiring finer sampling in the lens plane. Hence, we choose $n_{\rm gl}=1000$ for $w<10$, 2000 for $10<w<20$, \ldots, 10,000 for $90<w<100$ in both axisymmetric and general algorithms. The total number of nodes will be $N_{\text{gl,1D}} = n_{\text{gl}}$ and $N_{\text{gl,2D}} = n_{\text{gl}}^2$ in the axisymmetric and general cases, respectively. For each frequency (or frequency bin), we adaptively choose the integration window size $R = \sqrt{n_{\text{gl}}/(2w_{\text{max}})}$, where $w_{\text{max}}$ is either the current frequency of the largest frequency in the bin. This ensures that the quadratic Fresnel phase is adequately resolved. We impose a minimum floor on $R$ that is much larger than the physical lens scale and window our potentials as $\psi(x) \to \psi(x)W(x)$, with $W(x) = \left[1-\tanh(x-3 R/4) \right]/2$. Since the construction of GL quadratures is computationally expensive, the nodes and weights are precomputed and stored once per frequency bin, and reused in any subsequent evaluation of the integral, independent of the lens potential.

In the general case, we utilize the batch mode of \texttt{FINUFFT}'s Type-1 (for uniform source position grids) and Type-3 (for individual source points) NUFFTs to perform all FFTs for a given value of $n_{\text{gl}}$ with one call to the vectorized NUFFT, and parallelize the NUFFT with different GL quadratures on multiple cores. On 112 CPU cores on a computing cluster, the calculation takes $\sim 6$ seconds using a $500 \times 500$ source-plane grid and 560 uniformly spaced points between $w = 0.1$ and $w = 10$ for simple multi-component lens potentials.

In the axisymmetric case, our C code translates the Julia NUFHT package\footnote{\url{github.com/pbeckman/FastHankelTransform.jl}} into C, adding a batch mode to allow multiple NUFHTs on the same grid to be done at once. We do all the FHTs (for both real and imaginary parts) for a given value of $n_{\rm gl}$ with one call to the vectorized NUFHT and then parallelize the NUFHTs with different $n_{\rm gl}$ so they can be done simultaneously on multiple cores.  With these settings, the entire calculation takes $\sim0.7$ CPU seconds on a MacBook M2, which, when split between 12 cores, takes $\sim0.14$ seconds of wall time. The computational time for each $w$ is roughly proportional to $n_{\rm gl}$, so the total time increases roughly as the square of the maximum frequency (if the spacing in $w$ remains uniform). Given the structure of the NUFHT, an increase in the number of $y$ values from 150 to 1500 increases the run time by about a factor of 2. No further economy is obtained, however, by reducing the number below 150 (in fact, it takes longer for fewer values of $y$).

The vectorization approach offers the additional advantage that derivatives of $F(w,y)$ with respect to model parameters can be obtained---using the NUFHT or NUFFT batch modes---with only incremental additional computational time.  For example, if the potential is written $\psi(x,{\bf s})$ in terms of some vector of parameters ${\bf s}$, so that the Fresnel integral is generalized to $F(w,y;{\bf s)}$ then the partial derivatives $\partial F(w,y,{\bf s})/\partial s_j$ can be written as in Eq.~(\ref{eqn:fht}) but replacing $(e^{-i\psi}-1)$ in the integrand in Eq~(\ref{eqn:I}) by $-i (\partial \psi/\partial s_j) e^{-i\psi}$ and similarly for the coefficients in Eq.~(\ref{eqn:Isum}).  The NUFHT structure for $F(w,y;{\bf s})$ and all the $\partial F(w,y;{\bf s})/\partial s_j$ are the same and can be done with one call to the batch-mode NUFHT.

\section{Code Performance}
\label{sec:code_perf}

\begin{figure}[!t]
    \centering
    \includegraphics[width=\linewidth]{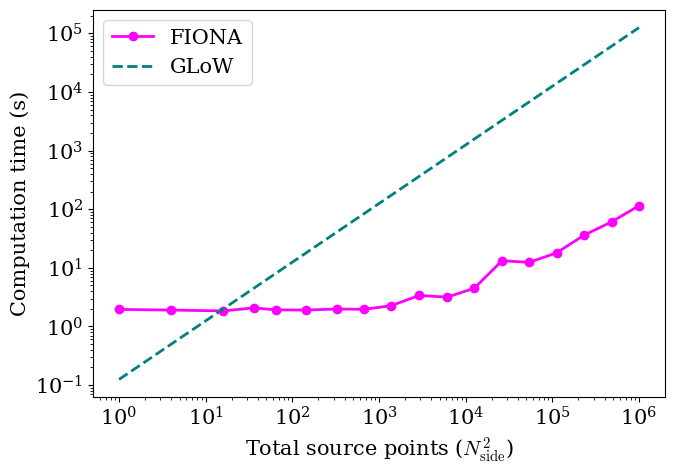}
    \caption{Scaling of computation time versus the total number of source points for the 1D Fresnel integral evaluated with \texttt{FIONA} and \texttt{GLoW} (\texttt{SingleIntegral\_C} function). The lens potential used for illustration is a singular isothermal sphere.}
    \label{fig:1d_scaling}
\end{figure}

\begin{figure}[!t]
    \centering
    \includegraphics[width=\linewidth]{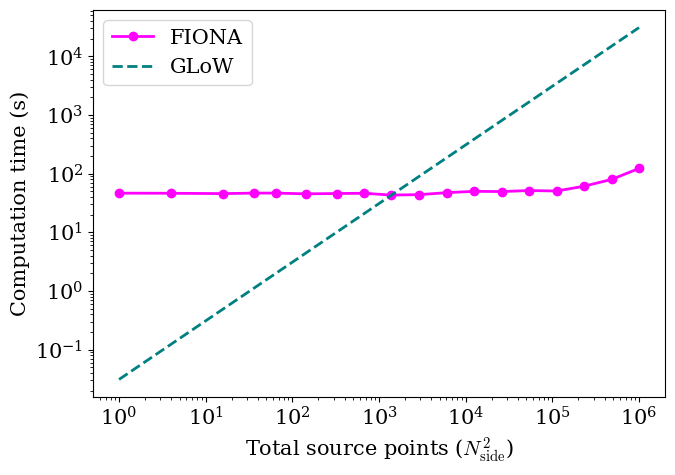}
    \caption{Scaling of computation time versus the total number of source points for the 2D Fresnel integral evaluated with \texttt{FIONA} and \texttt{GLoW} (\texttt{Multi\_Contour\_C} function). The lens potential is composed of 4 randomly-placed singular isothermal spheres.}
    \label{fig:2d_scaling}
\end{figure}

In this Section, we benchmark the performance of \texttt{FIONA} by comparing it to \texttt{GLoW} \cite{Villarrubia-Rojo:2024xcj}, a publicly available package for gravitational wave lensing computations. In Figs.~\ref{fig:1d_scaling} and \ref{fig:2d_scaling}, we show the scaling of computation with respect to the total number of source points for axisymmetric and general lens cases, respectively.

For the axisymmetric case (Fig.~\ref{fig:1d_scaling}), \texttt{FIONA} exhibits substantially shallower scaling with the number of source points $N_{\text{side}}^2$ than GLoW. The latter method is fast and efficient, but directly evaluates F(w) only for a single source point, leading to linear scaling with the number of source points. In contrast, \texttt{FIONA} remains nearly flat over several orders of magnitude in the number of source points, with a gradual increase at the largest values. Hence, \texttt{FIONA} outperforms \texttt{GLoW} for more than 10 source points, and is roughly 2 orders of magnitude faster for $100\times100$ source grids.

For the general case (Fig.~\ref{fig:2d_scaling}), \texttt{FIONA} maintains an almost constant runtime up to $10^5$ source points, with only a mild growth at the largest $N_{\text{side}}^2$. Hence, the code outperforms \texttt{GLoW} for dense source grids with more than $10^3$ points.

In Appendix ~A, we validate the accuracy of our code by performing cross-checks with GLoW.

\section{Representative Results}
\label{sec:results}

In this section, we demonstrate our method using different lens models and show its ability to efficiently compute the amplification factor $F(w,\mathbf{y})$ over a densely sampled source plane and across a broad frequency range. We begin with simple axisymmetric lenses and then consider more complex configurations, including elliptical, sheared, and multi-component mass distributions. Finally, we explore the use of wave-optics effects as a probe of dark-matter substructure through simple host–subhalo configurations. The explicit forms of all lens potentials used in this section are given in Appendix~B.

\subsection{Cored Isothermal Sphere}

\begin{figure}
    \centering
    \includegraphics[width=\linewidth]{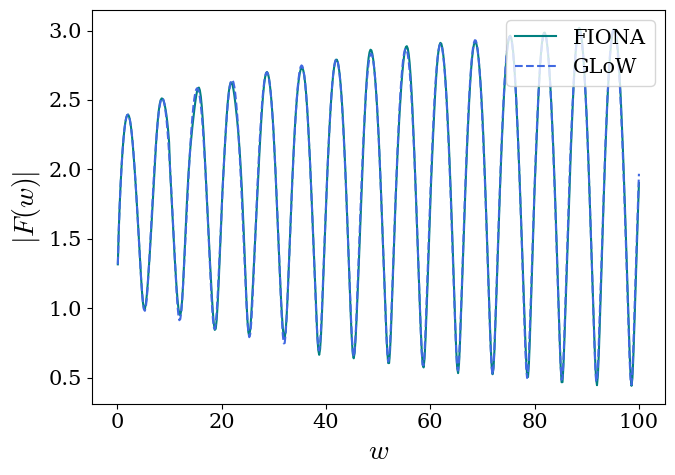}
    \caption{$F(w,y)$ for the cored isothermal sphere (CIS) with core radius $x_c = 0.05$ evaluated at $y=0.5$. The computation with \texttt{FIONA} takes 2.1 s for 560 uniformly spaced frequencies between $w=0.1$ and $w=100$ on 112 CPU cores. Results from \texttt{GLoW} are overplotted with a dashed line for validation.}
    \label{fig:Fy0.297}
\end{figure}

\begin{figure}
    \centering
    \includegraphics[width=\linewidth]{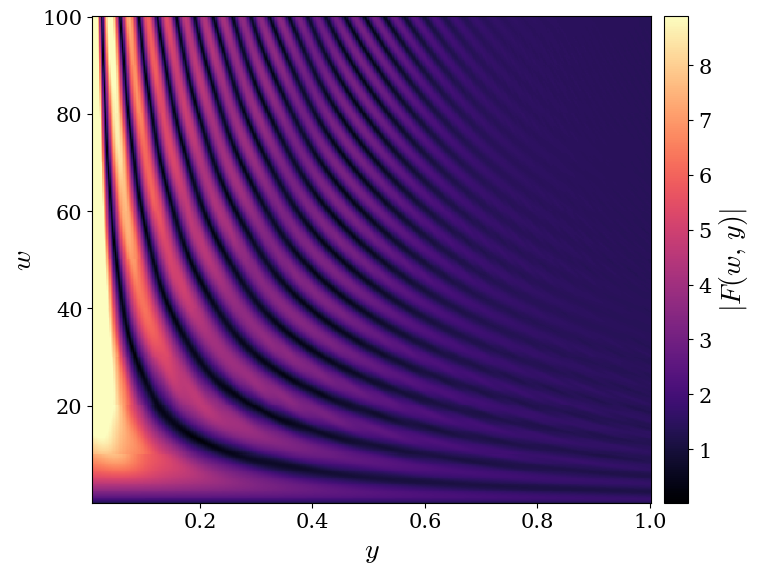}
    \caption{Contour plot $F(w,y)$ for a CIS lens with $x_c = 0.05$, inferred from 560 and 400 uniformly spaced points in $w$ and $y$, respectively. The computation takes 2.1 s on 112 CPU cores.}
    \label{fig:Fwy}
\end{figure}

We compute the amplification factor $F(w, y)$ for a cored isothermal sphere (CIS) with a core radius $x_c = 0.05$ using the NUFHT-based axisymmetric integral method (Section \ref{sec:axisym}). Figures~\ref{fig:Fy0.297} and ~\ref{fig:Fwy} show $F(w,y)$ evaluated at a fixed source position and over the $w$-$y$ plane. The computation in Figure~\ref{fig:Fy0.297} takes 2.1 s for 560 uniformly spaced frequencies between $0.1 \leq w \leq 100$ on 112 CPU cores. For the same frequencies but 400 different source positions, the computation (Figure~\ref{fig:Fwy}) takes the same time (2.1 s) on the same hardware. Once again, this illustrates the efficiency of an NUFHT-based approach for dense source-position grids, discussed in Section \ref{sec:code_perf}. Our method reliably captures the rapid oscillations that develop in $F(w,y)$ at high frequencies and large source offsets, as shown by excellent agreement between results from \texttt{FIONA} and \texttt{GLoW}.

\begin{figure*}[t]
    \centering
    \includegraphics[width=\textwidth]{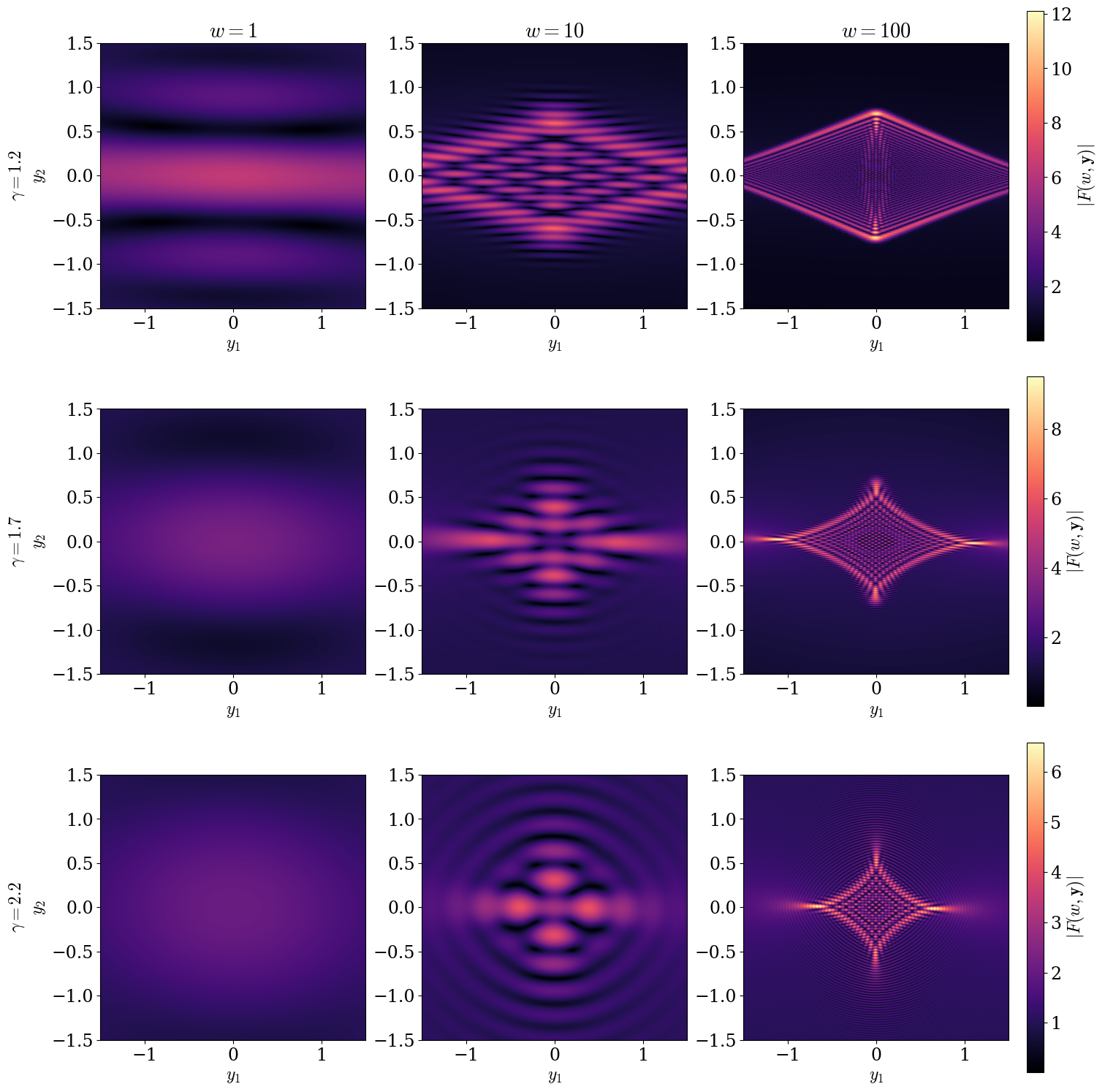}
    \caption{$|F(w, y)|$ for a lens with an Elliptical Power-Law profile ($e_1=0.2$, $e_2=0.0$) and a weak background shear ($\gamma_1$=0.03, $\gamma_2$=0.01), shown over the source plane $\mathbf{y} = (y_1, y_2)$ for varying negative density slope ($\gamma = 1.2, 1.7, 2.2$) at different frequencies ($w=1, 10, 100$). The computation of results for all three lenses, evaluated on a $500 \times 500$ source grid at the three selected frequencies, takes $\sim 100$ s on a single CPU core.}
    \label{fig:EPL_shear_grid}
\end{figure*}

\begin{figure*}[t]
    \centering
    \includegraphics[width=\textwidth]{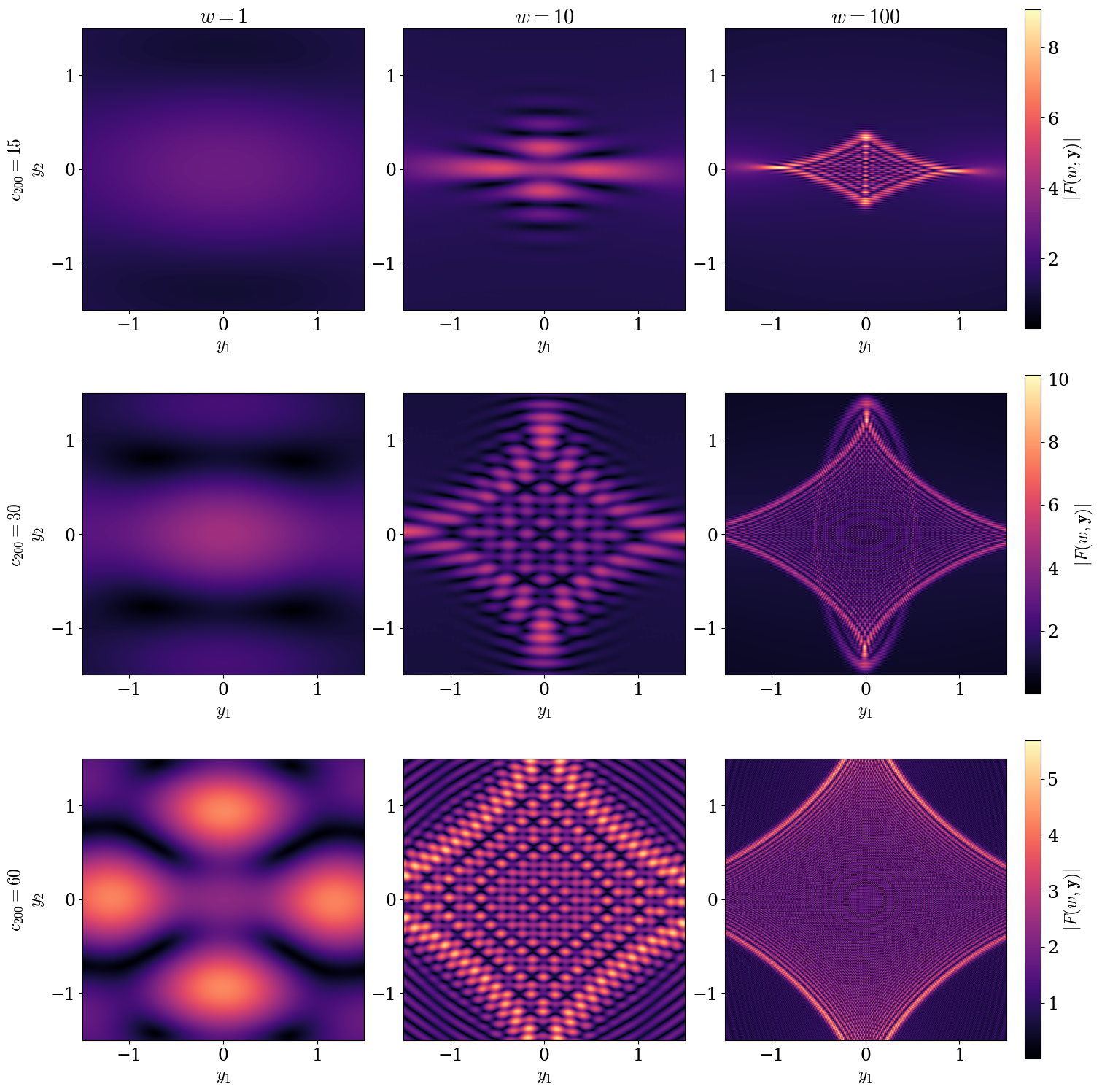}
    \caption{$|F(w, y)|$ for a lens with an elliptical Navarro-Frenk-White profile ($e_1=0.2$, $e_2=0.0$) and a weak background shear ($\gamma_1$=0.03, $\gamma_2$=0.01), shown over the source plane $\mathbf{y} = (y_1, y_2)$ for varying concentrations ($c_{200} = 15, 30, 60$) at different frequencies ($w=1, 10, 100$). The computation of results for all three lenses, evaluated on a $500 \times 500$ source grid at the three selected frequencies, takes $\sim 130$ s on a single CPU core.}
    \label{fig:eNFW_shear}
\end{figure*}

\subsection{Elliptical and Sheared Lenses}
We compute results for non-axisymmetric lenses using the general NUFFT-based method (Section~\ref{sec:general_lens}). Figure~\ref{fig:EPL_shear_grid} shows the amplitude of the amplification factor $|F(w,\mathbf{y})|$ for an elliptical power-law (EPL) lens with a small external shear and varying negative density slope, evaluated over the source plane at fixed frequencies $w=1,10,$ and $100$. The computation takes As the density slope $\gamma$ decreases, the extent of the interference pattern in the source plane increases, producing a larger and more strongly oscillating caustic structure at high frequency. At the highest frequency, $w=100$, the amplification pattern begins to approach the familiar diamond-shaped caustic morphology characteristic of the geometric-optics limit.

Figure~\ref{fig:eNFW_shear} presents analogous results for an elliptical Navarro--Frenk--White (NFW) halo with virial mass $M_{200}=10^{13}\,M_{\odot}$ and the same external shear, while varying the concentration $c_{200}$. As in the EPL case, increasing the concentration leads to a more rapidly oscillating structure of $F(\mathbf{y})$, and at high frequency the pattern approaches the geometric-optics caustic structure.

For the two different lens models (Figures ~\ref{fig:EPL_shear_grid} and ~\ref{fig:eNFW_shear}), the presented results are computed on a $500\times500$ source grid in $\sim100$ s and $\sim130$ s, respectively, using a single CPU core.

\begin{figure*}[htbp]
    \centering

    \begin{subfigure}[b]{\textwidth}
        \centering
        \includegraphics[width=\linewidth]{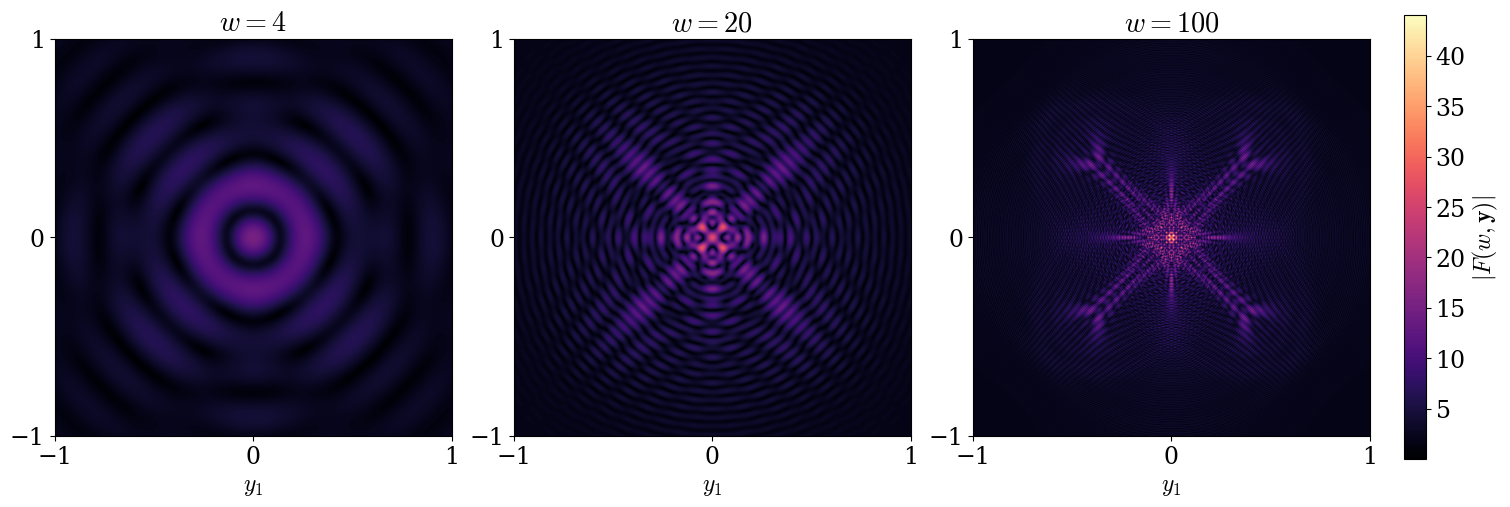}
        \caption{Magnitude of the amplification factor $|F(w,\mathbf{y})|$ over the source plane $\mathbf{y}=(y_1,y_2)$ at representative frequencies $w=4,\,20,$ and $100$.}
        \label{fig:4_CIS_F(y)}
    \end{subfigure}
    \hfill
    \begin{subfigure}[b]{\textwidth}
        \centering
        \includegraphics[width=\linewidth]{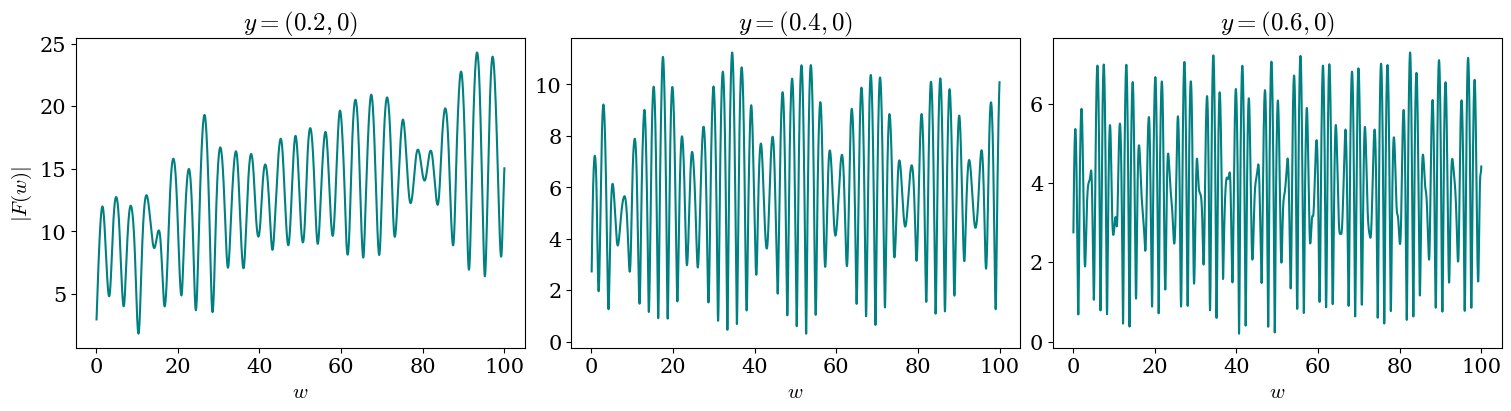}
        \caption{Frequency dependence of $|F(w)|$ evaluated at selected source positions.}
        \label{fig:4_CIS_Fw}
    \end{subfigure}

    \caption{
    Results for a lens composed of four cored isothermal spheres (CIS). 
    The amplification factor is computed using a $500\times500$ source-plane grid for the maps in panel (a), requiring 0.8 s, 6 s, and 75 s for $w=4$, 20, and 100, respectively, on a single CPU core. 
    The frequency-domain curves in panel (b) are computed using 560 uniformly spaced points between $w=0.1$ and $100$, requiring $181$ s on 112 CPU cores.
    }
    \label{fig:4_CIS}
\end{figure*}
\subsection{Multiple Lens Systems}

\begin{figure*}[htbp]
    \centering

    \begin{subfigure}[b]{\textwidth}
        \centering
        \includegraphics[width=\linewidth]{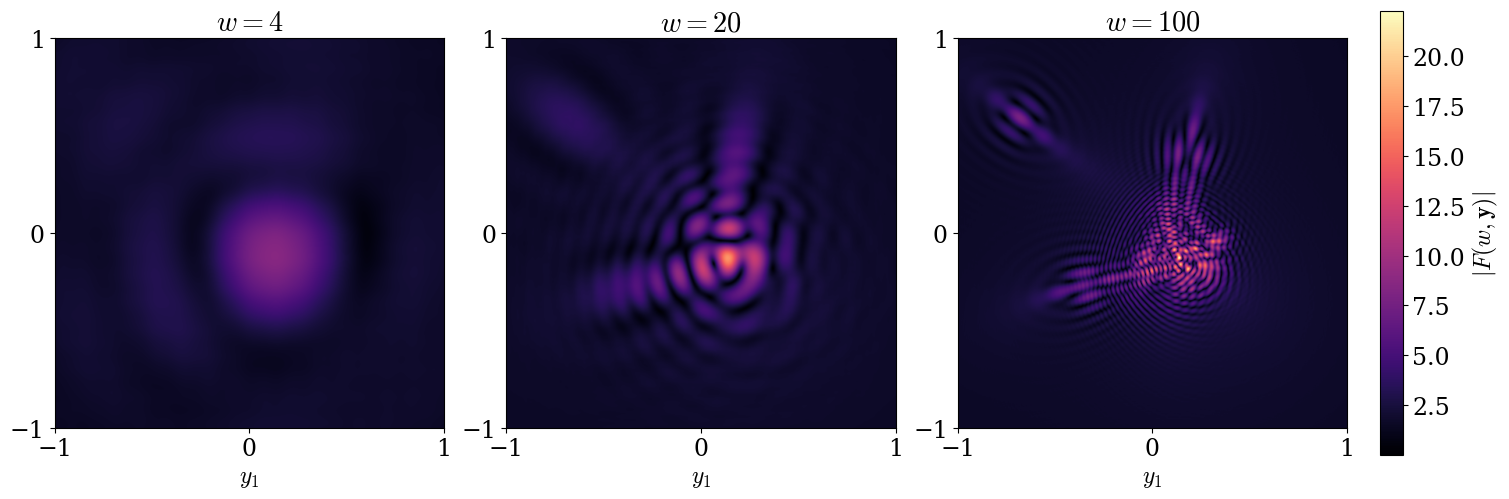}
        \caption{Magnitude of the amplification factor $|F(w,\mathbf{y})|$ over the source plane $\mathbf{y}=(y_1,y_2)$ at representative frequencies $w=1,\,10,$ and $100$.}
        \label{fig:10_SIS_F(y)}
    \end{subfigure}
    \hfill
     
    \begin{subfigure}[b]{\textwidth}
        \centering
        \includegraphics[width=\linewidth]{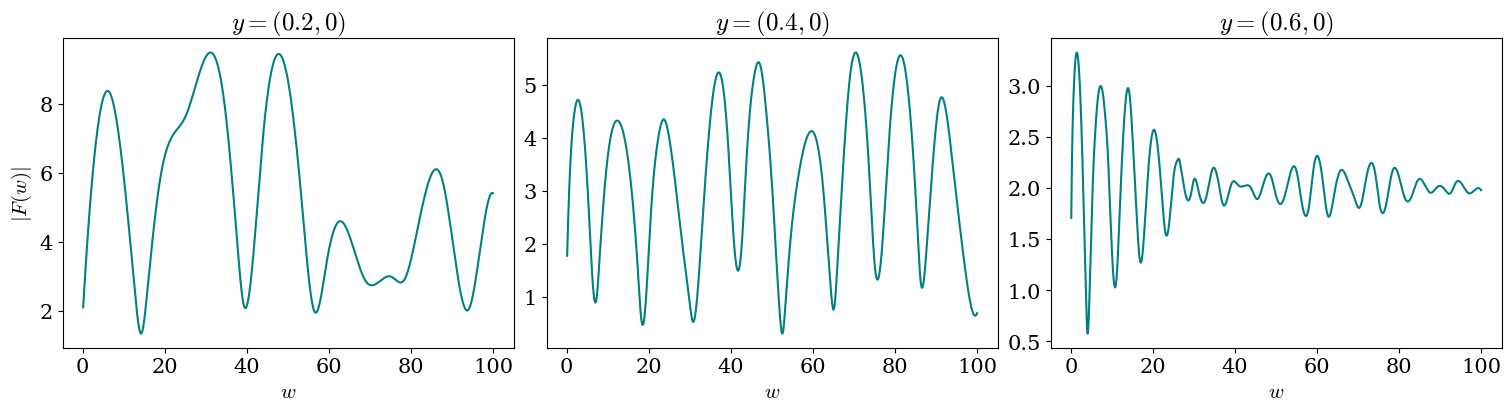}
        \caption{Frequency dependence of $|F(w)|$ evaluated at selected source positions.}
        \label{fig:10_SIS_Fw}
    \end{subfigure}
    
    \caption{
    Results for a lens composed of ten randomly placed singular isothermal spheres (SIS). 
    The source-plane maps in panel (a) are computed on a $500\times500$ grid, requiring 0.9 s, 6 s, and 84 s for $w=4$, 20, and 100, respectively, on a single CPU core. 
    The frequency-domain curves in panel (b) are computed using 560 uniformly spaced points between $w=0.1$ and $100$, requiring $205$ s on 112 CPU cores.
    }
    \label{fig:10_SIS}
\end{figure*}

We next consider composite lenses consisting of multiple mass components, illustrating the behavior of the method for increasingly complex interference structures.

Figure~\ref{fig:4_CIS} shows results for a lens composed of four cored isothermal spheres symmetrically arranged about the center. The source-plane maps in Fig.~\ref{fig:4_CIS_F(y)} display the amplification factor at $w=4$, 20, and 100 on a $500\times500$ grid. The computation requires 0.8 s, 6 s, and 75 s, respectively, on a single CPU core. As the frequency increases, the interference pattern becomes progressively more intricate, reflecting the superposition of multiple image contributions generated by the combined lens potential. The corresponding frequency-domain behavior at fixed source positions is shown in Fig.~\ref{fig:4_CIS_Fw}, computed using 560 uniformly spaced frequency points between $w=0.1$ and 100 in approximately 180 s on 112 CPU cores. The oscillatory modulation in $|F(w)|$ arises from interference between the multiple lensing paths of the system.

We next consider a more irregular configuration consisting of ten randomly placed singular isothermal spheres, shown in Fig.~\ref{fig:10_SIS}. The source-plane maps in Fig.~\ref{fig:10_SIS_F(y)} are computed on the same $500\times500$ grid, requiring 0.9 s, 6 s, and 84 s for $w=4$, 20, and 100, respectively, on a single core. The increased structural complexity of the lens produces correspondingly richer interference features. Interestingly, as $w$ increases, the phase of the Fresnel integral becomes increasingly sensitive to small-scale variations in the lensing potential, allowing individual clump-induced stationary points to be resolved and producing sharply localized, off-center caustic structures. The frequency-domain curves in Fig.~\ref{fig:10_SIS_Fw} are obtained using 560 frequency samples and require approximately 200 s on 112 CPU cores. Overall, these examples demonstrate that the Fourier-transform approach remains efficient and accurate even for multi-component, strongly non-axisymmetric lens configurations.

\subsection{Probing Dark Matter Substructure}

\begin{figure*}[p]
     \centering

     \begin{subfigure}[b]{\textwidth}
         \centering
         \includegraphics[width=\linewidth]{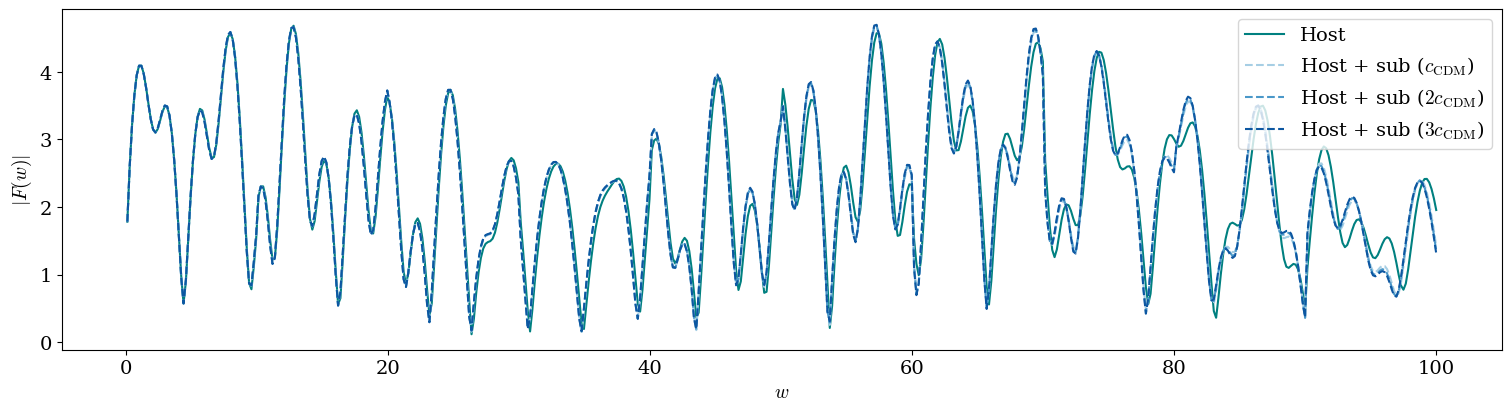}
         \caption{Subhalo mass $M_{\text{sub}}=10^9\,M_{\odot}$.}
         \label{fig:1e9_subhalo}
     \end{subfigure}
     \hfill
     
     \begin{subfigure}[b]{\textwidth}
         \centering
         \includegraphics[width=\linewidth]{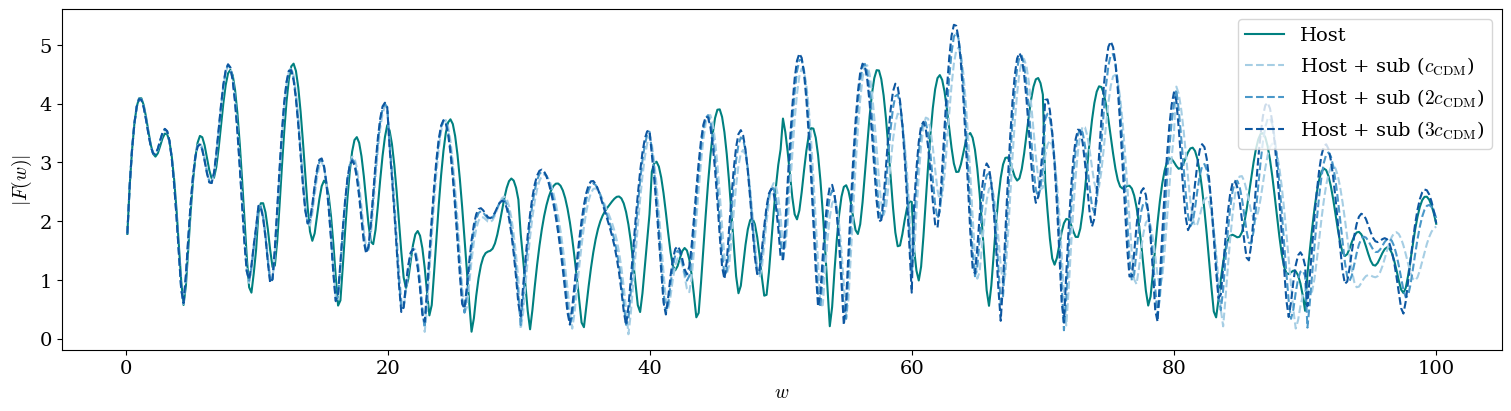}
         \caption{Subhalo mass $M_{\text{sub}}=10^{10}\,M_{\odot}$.}
         \label{fig:1e10_subhalo}
     \end{subfigure}
     \hfill
     
     \begin{subfigure}[b]{\textwidth}
         \centering
         \includegraphics[width=\linewidth]{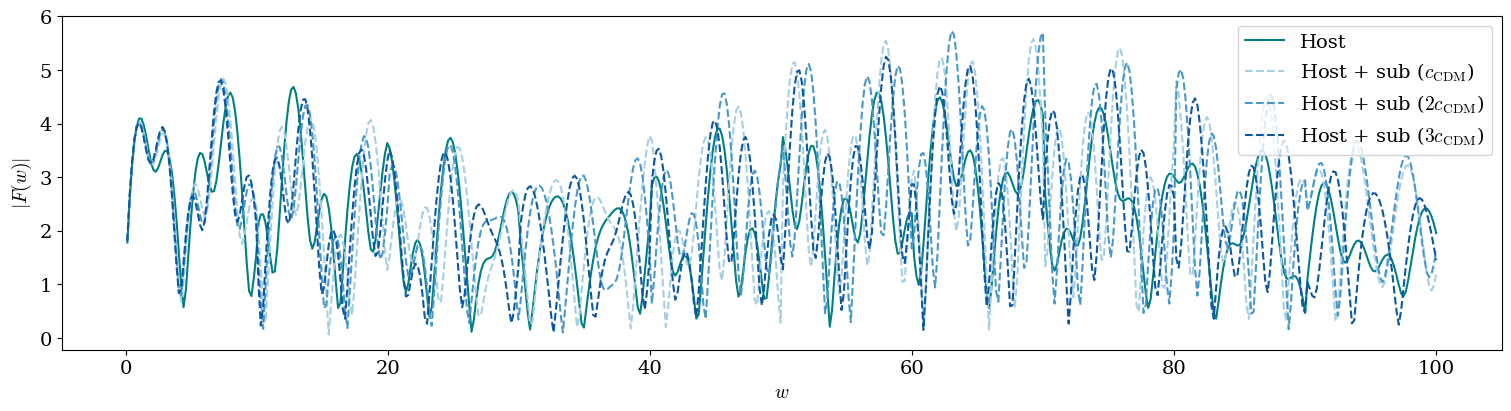}
         \caption{Subhalo mass $M_{\text{sub}}=10^{11}\,M_{\odot}$.}
         \label{fig:1e11_subhalo}
     \end{subfigure}
     
     \caption{
     Frequency-dependent amplification factor $F(w)$ for an elliptical NFW host halo with small external shear (same host model as in Fig.~\ref{fig:eNFW_shear}), including a spherical NFW subhalo of varying mass. The lens is at redshift $z_l=0.5$ and the source at $z_s=1.5$. The solid curve shows the host halo alone, while dashed curves include the subhalo contribution. For each subhalo mass, the dashed curves correspond to concentrations equal to 1, 2, and 3 times the value predicted by the standard cold dark matter mass–concentration relation. The subhalo is placed near the Einstein radius of the host, and $F(w,\mathbf{y})$ is evaluated at a source position chosen to maximize the perturbation induced by the subhalo.
     }
     \label{fig:Fw_subhalo}
\end{figure*}

GW lensing is emerging as a promising probe of dark matter (DM) substructure, with the potential to detect low-mass subhalos through their wave-optics signatures \cite{Oguri:2020ldf,Guo:2022dre,Caliskan:2023zqm,Cheung:2024ugg,Vujeva:2025xxx,Shinichiro:2026hfj} and thereby shed light on DM microphysics. As a simple toy model of this scenario, we consider a dark matter subhalo embedded within a galactic host halo. We model the host as an elliptical NFW profile with virial mass $M_{200}=10^{13}\,M_{\odot}$, concentration $c_{200}=10$, ellipticity parameters $(e_1,e_2)=(0.2,0)$, and a small external shear $(\gamma_1,\gamma_2)=(0.03,0.01)$ and place it at redshift $z_l=0.5$, with a GW source at $z_s=1.5$. We place a spherical NFW perturber with mass $M_{\text{sub}}=10^9$, $10^{10}$, or $10^{11}\,M_{\odot}$ and concentration equal to one, two, or three times the value predicted by the standard cold dark matter mass–concentration relation \cite{Bullock:2001nr,Diemer:2014xca} near the Einstein radius of the host. We evaluate the amplification factor at a source location that maximizes the lensing deflection induced by the perturber.

The resulting frequency-dependent amplification factor $F(w)$ is shown in Fig.~\ref{fig:Fw_subhalo}, with the three subhalo masses displayed in Figs.~\ref{fig:1e9_subhalo}, \ref{fig:1e10_subhalo}, and \ref{fig:1e11_subhalo}. The solid curve corresponds to the host halo alone, while the dashed curves include the additional contribution of the perturber.

The wave-optics effect of the perturber depends strongly on its mass, and to some extent, on its concentration. For $M_{\text{sub}}=10^9\,M_{\odot}$ (Fig.~\ref{fig:1e9_subhalo}), the host-only and host + subhalo $F(w)$ curves show only small deviations that only become visible toward higher frequencies ($w > 50$). This is to be expected, since the decreasing GW wavelength is able to probe a smaller characteristic physical scale of the subhalo. However, the curves with varying subhalo concentrations are essentially indistinguishable from one another. Increasing the subhalo mass to $10^{10}\,M_{\odot}$ (Fig.~\ref{fig:1e10_subhalo}) leads to visibly stronger wave-optics effects from the subhalo, including both phase shifts and enhanced peak amplitudes, with higher concentrations slightly amplifying the strength of the lensing effect. Finally, for a massive $M_{\text{sub}}=10^{11}\,M_{\odot}$ subhalo (Fig.~\ref{fig:1e11_subhalo}), both the phase and amplitude of $F(w)$ are significantly distorted relative to the host-only case, and the different concentration scalings are clearly separated for most of the frequency range.

\section{Conclusions}
\label{sec:conclusions}

Computing observables in GW lensing requires the evaluation of a Fresnel integral that appears in the amplification factor $F(w,\mathbf{y})$. Accurately evaluating the highly oscillatory integral for a broad range of frequencies, especially for complex lens potentials without spherical symmetry, is a computational bottleneck in GW lensing studies. In this work, we showed that the dependence of the Fresnel integral on the source position $\mathbf{y}$ can be written as a two-dimensional Fourier transform, which enables us to compute the integral with modern fast Fourier techniques. Using Gauss-Legendre quadratures for numerical integration and non-uniform FFT—or non-uniform fast Hankel transform for axisymmetric lenses—our method can compute $F(w,\mathbf{y})$ over a densely sampled source plane simultaneously for all source positions. Our results demonstrate that the algorithm is efficient and accurate for intermediate frequencies ($0.01 < w < 100$) across lens profiles ranging from axisymmetric lenses, elliptical and sheared lenses, to asymmetric and multi-component mass distributions.

Our approach offers key advantages to existing methods in GW lensing. The contour method \cite{Villarrubia-Rojo:2024xcj} is highly efficient at evaluating time-domain version of the amplification factor, which is then Fourier transformed to obtain $F(w)$. However, the evaluation is performed at a single source position $\mathbf{y}$, and the computation time grows proportionally to the number of points on the source grid. On the other hand, $\texttt{FIONA}$ evaluates all source positions simultaneously without additional computational cost (up to large i.e. $500 \times 500$ grids), significantly outperforming \cite{Villarrubia-Rojo:2024xcj} in evaluation of $F(w, \mathbf{y})$ over a dense grid of source points. Our approach is complementary to the Picard-Lefschetz method \cite{Feldbrugge:2019fjs, Feldbrugge:2020ycp}, which is well-suited for rapidly oscillating integrands and thereby efficient for large frequencies, but becomes less efficient at lower (intermediate) frequencies where $\texttt{FIONA}$ performs optimally.

The method developed here can make important contributions to the study of wave-optics phenomena in GW-lensing, which are emerging as novel probes of dark matter. The detection of gravitational waves, while some time in the future, is expected in upcoming observing surveys \cite{Oguri:2018mnr, Wierda:2021apj, Xu:2021bfn, Smith:2023bjh}. Analysis of lensed GW data will require exploring a high-dimensional parameter space to fit the signal model, where the position of the GW source is a key parameter (since the source is not resolved on the sky). Our method directly facilitates this analysis, as it allows lensed waveforms to be computed simultaneously for a large number of source positions. At the same time, our method is well-suited for batching and vectorization, allowing observables for multiple lens parameters and/or derivatives with respect to the model parameters to be computed with only incremental additional cost. This feature is useful not only for the above-mentioned Bayesian inference, but also for rapid Fisher forecasts to assess the constraining power of current and upcoming GW facilities on different dark matter theories.

Another interesting, although more futuristic, opportunity enabled by a computationally efficient access to $F(w,\mathbf{y})$ over the source plane is probing dark matter substructure in the transverse line of sight using time-domain information. If a lensed GW source exhibits detectable transverse motion relative to the lens (or if relative motion among observer, lens, and source induces a significant drift across $\mathbf{y}$), then at fixed GW frequency one could, in principle, observe temporal variations in the amplification as the system sweeps across the interference pattern in the source plane (e.g. see Figures \ref{fig:EPL_shear_grid} and \ref{fig:eNFW_shear}). The small-scale distribution of dark matter will determine the variation in the diffraction pattern across source positions, and thereby the variation of $F(w, \mathbf{y})$ in time, serving as a complementary dark matter probe. Such studies are still a long way off in the future, as they require angular localization and potentially a very long-duration monitoring a GW signal, but may nevertheless provide an interesting prospect for probing dark matter distribution with GW-lensing.

There are several directions for future work, including further developing the method and optimizing its implementation, as well as applying it to dark matter studies. Firstly, we have chosen to use Gauss-Legendre quadratures to numerically evaluate the integrals involved in GW lensing. While we have found that the approach is efficient and accurate, the quadrature may not be an optimal choice for our application due to the highly oscillatory nature of the integrands. Thus, it may be interesting to investigate the performance of numerical quadratures specifically designed for oscillatory functions. Secondly, we have shown that significant speedups can be gained if the lens is axisymmetric, using an approach based on one dimensional non-uniform fast Hankel transform. It may be possible to speed up computations for quadrupolar lenses (i.e. $\psi(\mathbf{x}) = \psi_0 + \psi_2(x)\cos{2\theta_{x}}$, where $\theta_{x}$ is the angle $\bf{x}$ makes with the $x_1$ axis) using a similar approach. Here, the single Bessel function from the axisymmetric case is replaced by an infinite series of Bessel functions that can be truncated around $k \sim w \psi_2(x)$. Thus, instead of non-uniform FFTs, one can use some number of non-uniform FHTs, which may be more efficient for small quadrupoles (i.e. nearly symmetric lenses with a small shear). Finally, while we make publicly available basic implementations of our algorithm in C and Python, it may be possible to further optimize our packages with additional effort. In terms of further applications, the developments in the computational aspects of GW-lensing opens an avenue for exploring wave-optics signatures of more complex lens configurations. Thus, it would be interesting to apply the method developed in this work to compute GW-lensing effects expected from the fiducial cold dark matter model and its alternatives. To achieve this, we must accurately and efficiently compute $F(w,\mathbf{y})$ for realistic configurations of full subhalo populations within the host and along the line of sight, simulated under different dark matter paradigms. It would be interesting to consider how alternative dark matter theories—such as warm, fuzzy, or self-interacting dark matter—that alter subhalo density profiles and their population statistics might modify the GW-lensing signals, offering an independent window into the nature of dark matter.

\begin{acknowledgments}
MK acknowledges useful discussions with Alex Barnett.
This work was supported at JHU by NSF Grant No.\ 2412361, NASA ATP Grant No.\ 80NSSC24K1226, and the Templeton Foundation. CD and NE were partially supported by the Department of Energy (DOE) Grant No. DE-SC0025671. 
\end{acknowledgments}

\appendix
\section{Comparison to Existing Methods}

\begin{figure}
    \centering
    \includegraphics[width=\linewidth]{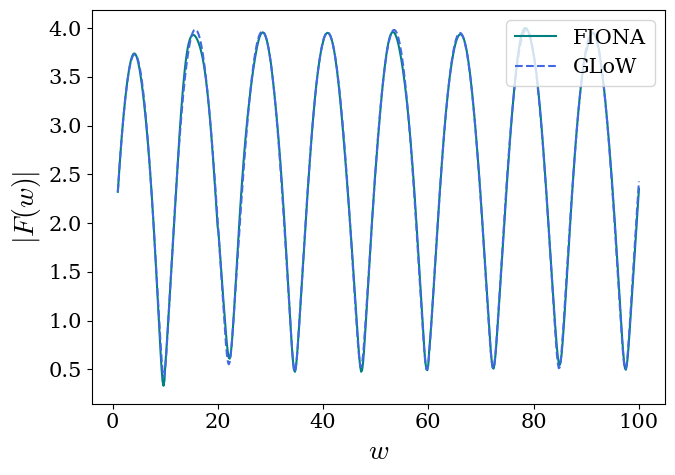}
    \caption{Comparison of $|F(w)|$ for an SIS lens ($\psi_0=1$) evaluated at source position $y=0.25$ using \texttt{FIONA} and \texttt{GLoW}.}
    \label{fig:axisym_GLoW_vs_FIONA}
\end{figure}

\begin{figure}
     \centering

     \begin{subfigure}[b]{\linewidth}
         \centering
         \includegraphics[width=\linewidth]{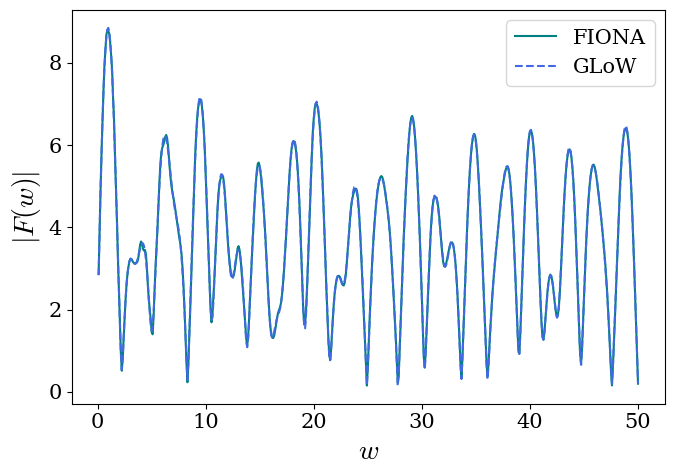}
         \caption{$y=0.25$.}
         \label{fig:4_SIS_GLoW_vs_FIONA_y=0.25}
     \end{subfigure}
     \hfill
     
     \begin{subfigure}[b]{\linewidth}
         \centering
         \includegraphics[width=\linewidth]{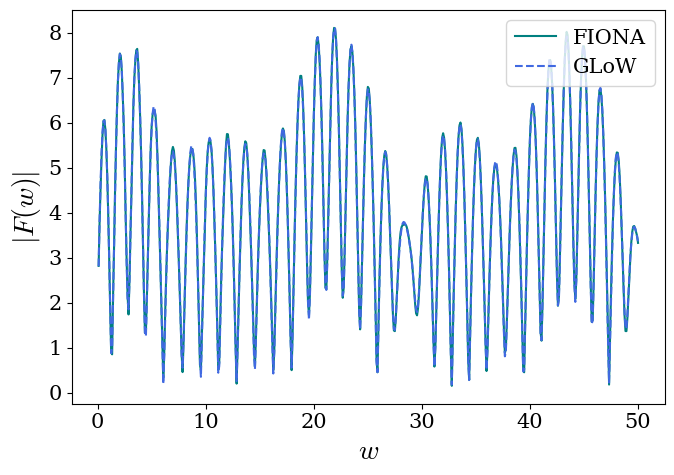}
         \caption{$y=0.5$.}
         \label{fig:4_SIS_GLoW_vs_FIONA_y=0.5}
     \end{subfigure}
     \hfill
     
     \begin{subfigure}[b]{\linewidth}
         \centering
         \includegraphics[width=\linewidth]{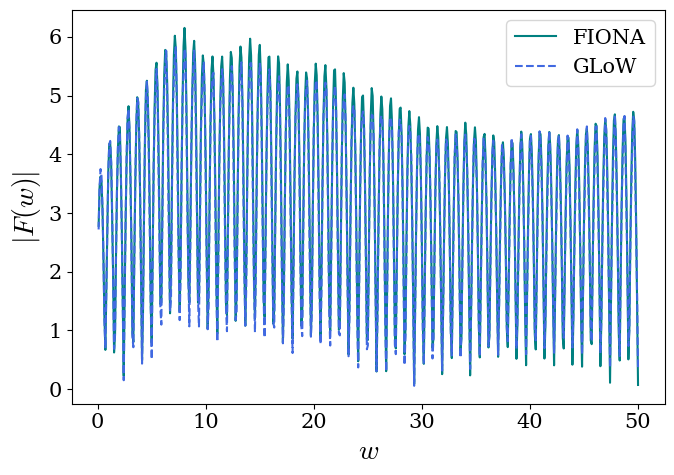}
         \caption{$y=1.0$.}
         \label{fig:4_SIS_GLoW_vs_FIONA_y=1.0}
     \end{subfigure}
     
     \caption{
     Comparison of results from \texttt{FIONA} and \texttt{GLoW} for a lens potential with 4 randomly-placed SIS profiles evaluated at different source positions.
     }
     \label{fig:4_SIS_GLoW_vs_FIONA}
\end{figure}

To validate the accuracy of our results and assess our code's performance, we perform cross-checks with the publicly available GW lensing package \texttt{GLoW} \cite{Villarrubia-Rojo:2024xcj}. In Figures \ref{fig:axisym_GLoW_vs_FIONA} and \ref{fig:4_SIS_GLoW_vs_FIONA}, we compare the results for the magnitude of the amplification factor $F(w)$ computed using \texttt{FIONA} and \texttt{GLoW} for axisymmetric (SIS profile) and general (4 SIS profiles) lenses, respectively. We find excellent agreement between the two methods.

\section{Lens Models}

In this Appendix, we summarize the lens potentials implemented in \texttt{FIONA} and used in demonstrations in this work.
Throughout, $\mathbf{x}=(x_1,x_2)$ denotes dimensionless coordinates in the lens plane
and $r \equiv |\mathbf{x}|$.

\subsection{Simple Axisymmetric Lenses}

\begin{itemize}

\item Singular isothermal sphere (SIS) \cite{Kormann:1994fz,Takahashi:2003ix}:
\begin{equation}
\psi_{\rm SIS}(r) = \psi_0\, r.
\end{equation}

\item Cored isothermal sphere (CIS):
\begin{equation}
\psi_{\rm CIS}(r)
=
\psi_0 \sqrt{r^2 + x_c^2}
+
\psi_0 x_c 
\ln\!\left[
\frac{2x_c}{\sqrt{r^2 + x_c^2}+x_c}
\right],
\end{equation}
which reduces to the SIS as the core radius $x_c \rightarrow 0$.

\item Plummer-softened point mass:
\begin{equation}
\psi_{\rm PM}(r)
=
\frac{1}{2}\psi_0 \ln(r^2 + x_c^2).
\end{equation}

\end{itemize}

\subsection{External Shear}

We model an external tidal shear field as
\begin{equation}
\psi_{\rm sh}(x_1,x_2)
=
\frac{1}{2}\gamma_1 (x_1^2 - x_2^2)
+
\gamma_2 x_1 x_2,
\end{equation}
where $(\gamma_1,\gamma_2)$ are the Cartesian shear components.

\subsection{Elliptical Power-Law (EPL)}

An EPL potential \cite{Kormann:1994fz,Barkana:1998} is characterized by a
three-dimensional power-law density slope $\gamma$. In the circular case, the potential may be written in terms of a generalized radial
coordinate $p$ as
\begin{equation}
\psi_{\rm EPL}
=
\frac{2E^2}{\eta^2}
\left(
\frac{p^2}{E^2}
\right)^{\eta/2},
\qquad
\eta = 3 - \gamma,
\end{equation}
where $E$ is a normalization related to the effective Einstein
radius $\theta_E$.

We introduce ellipticity by
replacing the circular radius with an elliptical radius aligned
with the principal axes. The complex ellipticity components
$(e_1,e_2)$ determine the position angle $\phi$ and axis ratio $q$
through
\begin{equation}
\phi = \frac{1}{2}\arctan2(e_2,e_1),
\qquad
q = \frac{1 - \sqrt{e_1^2+e_2^2}}{1 + \sqrt{e_1^2+e_2^2}}.
\end{equation}
Defining rotated coordinates
\begin{align}
x_t &= \cos\phi\,(x_1-x_c) + \sin\phi\,(x_2-y_c), \\
y_t &= -\sin\phi\,(x_1-x_c) + \cos\phi\,(x_2-y_c),
\end{align}
the elliptical radius is
\begin{equation}
p^2 = x_t^2 + \frac{y_t^2}{q^2}.
\end{equation}
The normalization is given by
\begin{equation}
E
=
\frac{\theta_E q}
{
\left(\frac{3-\gamma}{2}\right)^{1/(1-\gamma)}
\sqrt{q}
}.
\end{equation}
This construction follows the implementation adopted in the
\texttt{lenstronomy} package \cite{Birrer:2015xua,Birrer:2018tcu}.

\subsection{Navarro--Frenk--White (NFW)}

The spherical NFW \cite{Navarro:1996gj,Navarro:1997ks} potential is written as
\begin{equation}
\psi_{\rm NFW}(r)
=
\frac{1}{2}\psi_0
\left[
\ln^2\!\left(\frac{u}{2}\right)
+
(u^2 - 1)F(u)^2
\right],
\qquad
u = \frac{r}{x_s},
\end{equation}
where $x_s$ is the scale radius and
\begin{equation}
F(u)
=
\begin{cases}
\dfrac{1}{\sqrt{u^2 - 1}}
\arctan\!\left(\sqrt{u^2 - 1}\right), & u>1, \\[8pt]
\dfrac{1}{\sqrt{1-u^2}}
\operatorname{arctanh}\!\left(\sqrt{1-u^2}\right), & u<1, \\[8pt]
1, & u=1.
\end{cases}
\end{equation}

We construct an elliptical NFW potential following the implementation adopted in
\texttt{lenstronomy} \cite{Birrer:2015xua,Birrer:2018tcu}. The parameters
$(e_1,e_2)$ determine the position angle $\phi$ and axis ratio $q$ as
described above. After rotating into principal axes $(x_t,y_t)$,
one defines
\begin{equation}
e = \frac{|1-q^2|}{1+q^2},
\end{equation}
and maps the coordinates according to
\begin{align}
x' &= \sqrt{1-e}\,x_t, \\
y' &= \sqrt{1+e}\,y_t.
\end{align}
The spherical NFW potential is then evaluated at
\begin{equation}
R = \sqrt{x'^2 + y'^2},
\end{equation}
so that the elliptical potential is
\begin{equation}
\psi_{\rm eNFW}(x_1,x_2)
=
\psi_{\rm NFW}(R).
\end{equation}

\bibliography{refs}

@ARTICLE{Shinichiro:2026hfj,
       author = {{Ando}, Shin'ichiro},
        title = "{Wave-Optics Imprints of Dark Matter Subhalos on Strongly Lensed Gravitational Waves}",
      journal = {arXiv e-prints},
         year = 2026,
        month = mar,
          eid = {arXiv:2603.04267},
        pages = {arXiv:2603.04267},
archivePrefix = {arXiv},
       eprint = {2603.04267},
 primaryClass = {astro-ph.CO},
       adsurl = {https://ui.adsabs.harvard.edu/abs/2026arXiv260304267A}
}

@ARTICLE{Oguri:2018mnr,
       author = {{Oguri}, Masamune},
        title = "{Effect of gravitational lensing on the distribution of gravitational waves from distant binary black hole mergers}",
      journal = {Monthly Notices of the Royal Astronomical Society},
         year = 2018,
        month = nov,
       volume = {480},
       number = {3},
        pages = {3842-3855},
          doi = {10.1093/mnras/sty2145},
archivePrefix = {arXiv},
       eprint = {1807.02584},
 primaryClass = {astro-ph.CO},
       adsurl = {https://ui.adsabs.harvard.edu/abs/2018MNRAS.480.3842O}
}

@ARTICLE{Wierda:2021apj,
       author = {{Wierda}, A. Renske A.~C. and {Wempe}, Ewoud and {Hannuksela}, Otto A. and {Koopmans}, L{\'e}on V.~E. and {Van Den Broeck}, Chris},
        title = "{Beyond the Detector Horizon: Forecasting Gravitational-Wave Strong Lensing}",
      journal = {Astrophysical Journal},
         year = 2021,
        month = nov,
       volume = {921},
       number = {2},
          eid = {154},
        pages = {154},
          doi = {10.3847/1538-4357/ac1bb4},
archivePrefix = {arXiv},
       eprint = {2106.06303},
 primaryClass = {astro-ph.HE},
       adsurl = {https://ui.adsabs.harvard.edu/abs/2021ApJ...921..154W}
}

@article{Smith:2023bjh,
    author = {Smith, Graham P and Robertson, Andrew and Mahler, Guillaume and Nicholl, Matt and Ryczanowski, Dan and Bianconi, Matteo and Sharon, Keren and Massey, Richard and Richard, Johan and Jauzac, Mathilde},
    title = {Discovering gravitationally lensed gravitational waves: predicted rates, candidate selection, and localization with the Vera Rubin Observatory},
    journal = {Monthly Notices of the Royal Astronomical Society},
    volume = {520},
    number = {1},
    pages = {702-721},
    year = {2023},
    month = {01},
    issn = {0035-8711},
    doi = {10.1093/mnras/stad140},
    url = {https://doi.org/10.1093/mnras/stad140},
}

@article{Bullock:2001nr,
  author = {Bullock, James S. and Kolatt, Tsafrir S. and Sigad, Yuval and Somerville, Rachel S. and Kravtsov, Andrey V. and Klypin, Anatoly A. and Primack, Joel R. and Dekel, Avishai},
  title = {Profiles of dark haloes: evolution, scatter and environment},
  journal = {Mon. Not. Roy. Astron. Soc.},
  volume = {321},
  pages = {559--575},
  year = {2001},
  eprint = {astro-ph/9908159},
  archivePrefix = {arXiv},
  primaryClass = {astro-ph}
}

@article{Diemer:2014xca,
  author = {Diemer, Benedikt and Kravtsov, Andrey V.},
  title = {A Universal Model for Halo Concentrations},
  journal = {Astrophys. J.},
  volume = {799},
  pages = {108},
  year = {2015},
  eprint = {1407.4730},
  archivePrefix = {arXiv},
  primaryClass = {astro-ph.CO}
}

@article{Vujeva:2025xxx,
    author       = {Vujeva, Luka and Ezquiaga, Jose Mar{\'\i}a and Gilman, Daniel and Goyal, Srashti and Zumalac{\'a}rregui, Miguel},
    title        = {Dark Matter Subhalos and Higher Order Catastrophes in Gravitational Wave Lensing},
    eprint       = {2510.14953},
    archivePrefix= {arXiv},
    primaryClass = {astro-ph.CO},
    year         = {2025},
    note         = {Submitted},
    journal = ""
}

@article{Birrer:2015xua,
  author       = {Birrer, Simon and Amara, Adam and Refregier, Alexandre},
  title        = {Gravitational lens modeling with lenstronomy},
  journal      = {Journal of Cosmology and Astroparticle Physics},
  volume       = {2016},
  number       = {08},
  pages        = {020},
  year         = {2016},
  eprint       = {1511.03671},
  archivePrefix= {arXiv},
  primaryClass = {astro-ph.IM}
}

@article{Birrer:2018tcu,
  author       = {Birrer, Simon and Amara, Adam},
  title        = {lenstronomy: Multi-purpose gravitational lens modeling software package},
  journal      = {Physics of the Dark Universe},
  volume       = {22},
  pages        = {189--201},
  year         = {2018},
  eprint       = {1803.09746},
  archivePrefix= {arXiv},
  primaryClass = {astro-ph.IM}
}

@article{Navarro:1996gj,
  author       = {Navarro, Julio F. and Frenk, Carlos S. and White, Simon D. M.},
  title        = {The Structure of Cold Dark Matter Halos},
  journal      = {Astrophysical Journal},
  volume       = {462},
  pages        = {563--575},
  year         = {1996},
  eprint       = {astro-ph/9508025},
  archivePrefix= {arXiv}
}

@article{Navarro:1997ks,
  author       = {Navarro, Julio F. and Frenk, Carlos S. and White, Simon D. M.},
  title        = {A Universal Density Profile from Hierarchical Clustering},
  journal      = {Astrophysical Journal},
  volume       = {490},
  pages        = {493--508},
  year         = {1997},
  eprint       = {astro-ph/9611107},
  archivePrefix= {arXiv}
}

@article{Kormann:1994fz,
  author       = {Kormann, Rainer and Schneider, Peter and Bartelmann, Matthias},
  title        = {Isothermal elliptical gravitational lens models},
  journal      = {Astronomy and Astrophysics},
  volume       = {284},
  pages        = {285--299},
  year         = {1994},
  eprint       = {astro-ph/9303002},
  archivePrefix= {arXiv}
}

@article{Barkana:1998,
  author = {Barkana, Rennan},
  title = {Fast calculation of a family of elliptical gravitational lens models},
  journal = {Astrophysical Journal},
  volume = {502},
  pages = {531--537},
  year = {1998},
  eprint = {astro-ph/9802002}
}

@article{Barnett2021,
  author = {Barnett, Alex H.},
  title = {Efficient high-order accurate Fresnel diffraction via areal quadrature and the nonuniform fast Fourier transform},
  journal = {Open Access Journal of Astronomical Telescopes, Instruments, and Systems},
  volume = {7},
  number = {2},
  pages = {21211},
  year = {2021},
  month = {Jan},
  doi = {},
}

@misc{finufft,
  howpublished = {\url{https://finufft.readthedocs.io/en/latest/}},
}

@article{Fairbairn:2022xln,
    author = "Fairbairn, Malcolm and Urrutia, Juan and Vaskonen, Ville",
    title = "{Microlensing of gravitational waves by dark matter structures}",
    eprint = "2210.13436",
    archivePrefix = "arXiv",
    primaryClass = "astro-ph.CO",
    doi = "10.1088/1475-7516/2023/07/007",
    journal = "JCAP",
    volume = "07",
    pages = "007",
    year = "2023"
}

@article{Tambalo:2022plm,
    author = "Tambalo, Giovanni and Zumalac{\'a}rregui, Miguel and Dai, Liang and Cheung, Mark Ho-Yeuk",
    title = "{Lensing of gravitational waves: Efficient wave-optics methods and validation with symmetric lenses}",
    eprint = "2210.05658",
    archivePrefix = "arXiv",
    primaryClass = "gr-qc",
    doi = "10.1103/PhysRevD.108.043527",
    journal = "Phys. Rev. D",
    volume = "108",
    number = "4",
    pages = "043527",
    year = "2023"
}

@article{Guo:2022dre,
    author = "Guo, Xiao and Lu, Youjun",
    title = "{Probing the nature of dark matter via gravitational waves lensed by small dark matter halos}",
    eprint = "2207.00325",
    archivePrefix = "arXiv",
    primaryClass = "astro-ph.CO",
    doi = "10.1103/PhysRevD.106.023018",
    journal = "Phys. Rev. D",
    volume = "106",
    number = "2",
    pages = "023018",
    year = "2022"
}

@article{Villarrubia-Rojo:2024xcj,
    author = "Villarrubia-Rojo, Hector and Savastano, Stefano and Zumalac{\'a}rregui, Miguel and Choi, Lyla and Goyal, Srashti and Dai, Liang and Tambalo, Giovanni",
    title = "{Gravitational lensing of waves: Novel methods for wave-optics phenomena}",
    eprint = "2409.04606",
    archivePrefix = "arXiv",
    primaryClass = "gr-qc",
    doi = "10.1103/PhysRevD.111.103539",
    journal = "Phys. Rev. D",
    volume = "111",
    number = "10",
    pages = "103539",
    year = "2025"
}

@article{Cheung:2024ugg,
    author = "Cheung, Mark Ho-Yeuk and Ng, Ken K. Y. and Zumalac{\'a}rregui, Miguel and Berti, Emanuele",
    title = "{Probing minihalo lenses with diffracted gravitational waves}",
    eprint = "2403.13876",
    archivePrefix = "arXiv",
    primaryClass = "gr-qc",
    doi = "10.1103/PhysRevD.109.124020",
    journal = "Phys. Rev. D",
    volume = "109",
    number = "12",
    pages = "124020",
    year = "2024"
}

@article{Caliskan:2023zqm,
    author = "{\c{C}}al{\i}{\c{s}}kan, Mesut and Anil Kumar, Neha and Ji, Lingyuan and Ezquiaga, Jose M. and Cotesta, Roberto and Berti, Emanuele and Kamionkowski, Marc",
    title = "{Probing wave-optics effects and low-mass dark matter halos with lensing of gravitational waves from massive black holes}",
    eprint = "2307.06990",
    archivePrefix = "arXiv",
    primaryClass = "astro-ph.CO",
    doi = "10.1103/PhysRevD.108.123543",
    journal = "Phys. Rev. D",
    volume = "108",
    number = "12",
    pages = "123543",
    year = "2023"
}

@article{Ulmer:1994ij,
    author = "Ulmer, Andrew and Goodman, Jeremy",
    title = "{Femtolensing: Beyond the semiclassical approximation}",
    eprint = "astro-ph/9406042",
    archivePrefix = "arXiv",
    doi = "10.1086/175422",
    journal = "Astrophys. J.",
    volume = "442",
    pages = "67",
    year = "1995"
}

@article{BarnettMaglandKlinteberg2019,
  author    = {Barnett, Alex H. and Magland, Jeremy F. and af Klinteberg, Ludvig},
  title     = {A parallel non-uniform fast Fourier transform library based on an ``exponential of semicircle'' kernel},
  journal   = {SIAM Journal on Scientific Computing},
  volume    = {41},
  number    = {5},
  pages     = {C479--C504},
  year      = {2019},
  eprint    = {1808.06736},
  doi       = {10.1137/18M120885X},
  url       = {https://doi.org/10.1137/18M120885X}
}

@article{Barnett2021a,
  author    = {Barnett, Alex H.},
  title     = {Aliasing error of the exp kernel in the nonuniform fast Fourier transform},
  journal   = {Applied and Computational Harmonic Analysis},
  volume    = {51},
  pages     = {1--16},
  year      = {2021},
  eprint    = {2001.09405},
  doi       = {10.1016/j.acha.2020.10.003},
  url       = {https://www.sciencedirect.com/science/article/pii/S1063520320300725}
}

@article{Feldbrugge:2019fjs,
    author = "Feldbrugge, Job and Pen, Ue-Li and Turok, Neil",
    title = "{Oscillatory path integrals for radio astronomy}",
    eprint = "1909.04632",
    archivePrefix = "arXiv",
    primaryClass = "astro-ph.HE",
    doi = "10.1016/j.aop.2023.169255",
    journal = "Annals Phys.",
    volume = "451",
    pages = "169255",
    year = "2023"
}

@article{Feldbrugge:2020ycp,
    author = "Feldbrugge, Job and Turok, Neil",
    title = "{Gravitational lensing of binary systems in wave optics}",
    eprint = "2008.01154",
    archivePrefix = "arXiv",
    primaryClass = "gr-qc",
    month = "8",
    year = "2020",
    journal = ""
}

@article{Grillo:2018qjt,
    author = "Grillo, Gianfranco and Cordes, James",
    title = "{Wave asymptotics and their application to astrophysical plasma lensing}",
    eprint = "1810.09058",
    archivePrefix = "arXiv",
    primaryClass = "astro-ph.CO",
    month = "10",
    year = "2018",
    journal = ""
}

@misc{beckman2024,
      title={A Nonuniform Fast Hankel Transform}, 
      author={Paul G. Beckman and Michael O'Neil},
      year={2024},
      eprint={2411.09583},
      archivePrefix={arXiv},
      primaryClass={math.NA},
      url={https://arxiv.org/abs/2411.09583}, 
}

@article{Hamilton:1999uv,
    author = "Hamilton, A. J. S.",
    title = "{Uncorrelated modes of the nonlinear power spectrum}",
    eprint = "astro-ph/9905191",
    archivePrefix = "arXiv",
    doi = "10.1046/j.1365-8711.2000.03071.x",
    journal = "Mon. Not. Roy. Astron. Soc.",
    volume = "312",
    pages = "257--284",
    year = "2000"
}

@article{Ohanian:1974ys,
    author = "Ohanian, H. C.",
    title = "{On the focusing of gravitational radiation}",
    doi = "10.1007/BF01810927",
    journal = "Int. J. Theor. Phys.",
    volume = "9",
    pages = "425--437",
    year = "1974"
}

@article{Deguchi:1986zz,
    author = "Deguchi, Shuji and Watson, William D.",
    title = "{Wave effects in gravitational lensing of electromagnetic radiation}",
    doi = "10.1103/PhysRevD.34.1708",
    journal = "Phys. Rev. D",
    volume = "34",
    pages = "1708--1718",
    year = "1986"
}

@article{Wang:1996as,
    author = "Wang, Yun and Stebbins, Albert and Turner, Edwin L.",
    title = "{Gravitational lensing of gravitational waves from merging neutron star binaries}",
    eprint = "astro-ph/9605140",
    archivePrefix = "arXiv",
    reportNumber = "FERMILAB-PUB-96-091-A",
    doi = "10.1103/PhysRevLett.77.2875",
    journal = "Phys. Rev. Lett.",
    volume = "77",
    pages = "2875--2878",
    year = "1996"
}

@article{Nakamura:1997sw,
    author = "Nakamura, Takahiro T.",
    title = "{Gravitational lensing of gravitational waves from inspiraling binaries by a point mass lens}",
    reportNumber = "UTAP-272-97, YITP-97-61",
    doi = "10.1103/PhysRevLett.80.1138",
    journal = "Phys. Rev. Lett.",
    volume = "80",
    pages = "1138--1141",
    year = "1998"
}

@article{Takahashi:2003ix,
    author = "Takahashi, Ryuichi and Nakamura, Takashi",
    title = "{Wave effects in gravitational lensing of gravitational waves from chirping binaries}",
    eprint = "astro-ph/0305055",
    archivePrefix = "arXiv",
    doi = "10.1086/377430",
    journal = "Astrophys. J.",
    volume = "595",
    pages = "1039--1051",
    year = "2003"
}

@article{Oguri:2020ldf,
    author = "Oguri, Masamune and Takahashi, Ryuichi",
    title = "{Probing Dark Low-mass Halos and Primordial Black Holes with Frequency-dependent Gravitational Lensing Dispersions of Gravitational Waves}",
    eprint = "2007.01936",
    archivePrefix = "arXiv",
    primaryClass = "astro-ph.CO",
    doi = "10.3847/1538-4357/abafab",
    journal = "Astrophys. J.",
    volume = "901",
    number = "1",
    pages = "58",
    year = "2020"
}

@article{Cremonese:2021puh,
    author = "Cremonese, Paolo and Ezquiaga, Jose Mar{\'\i}a and Salzano, Vincenzo",
    title = "{Breaking the mass-sheet degeneracy with gravitational wave interference in lensed events}",
    eprint = "2104.07055",
    archivePrefix = "arXiv",
    primaryClass = "astro-ph.CO",
    doi = "10.1103/PhysRevD.104.023503",
    journal = "Phys. Rev. D",
    volume = "104",
    number = "2",
    pages = "023503",
    year = "2021"
}

@article{Gao:2021sxw,
    author = "Gao, Zucheng and Chen, Xian and Hu, Yi-Ming and Zhang, Jian-Dong and Huang, Shun-Jia",
    title = "{A higher probability of detecting lensed supermassive black hole binaries by LISA}",
    eprint = "2102.10295",
    archivePrefix = "arXiv",
    primaryClass = "astro-ph.CO",
    doi = "10.1093/mnras/stac365",
    journal = "Mon. Not. Roy. Astron. Soc.",
    volume = "512",
    number = "1",
    pages = "1--10",
    year = "2022"
}

@article{DePaolis:2002tw,
    author = "De Paolis, Francesco and Ingrosso, G. and Nucita, A. A. and Qadir, A.",
    editor = "Gorhamn, Peter W.",
    title = "{A Note on gravitational wave lensing}",
    eprint = "astro-ph/0209149",
    archivePrefix = "arXiv",
    reportNumber = "UNILE-1090902",
    doi = "10.1051/0004-6361:20021258",
    journal = "Astron. Astrophys.",
    volume = "394",
    pages = "749--752",
    year = "2002"
}

@article{Meena:2019ate,
    author = "Meena, Ashish Kumar and Bagla, J. S.",
    title = "{Gravitational lensing of gravitational waves: wave nature and prospects for detection}",
    eprint = "1903.11809",
    archivePrefix = "arXiv",
    primaryClass = "astro-ph.CO",
    doi = "10.1093/mnras/stz3509",
    journal = "Mon. Not. Roy. Astron. Soc.",
    volume = "492",
    number = "1",
    pages = "1127--1134",
    year = "2020"
}

@article{Sereno:2010dr,
    author = "Sereno, M. and Sesana, A. and Bleuler, A. and Jetzer, Ph. and Volonteri, M. and Begelman, M. C.",
    title = "{Strong lensing of gravitational waves as seen by LISA}",
    eprint = "1011.5238",
    archivePrefix = "arXiv",
    primaryClass = "astro-ph.CO",
    doi = "10.1103/PhysRevLett.105.251101",
    journal = "Phys. Rev. Lett.",
    volume = "105",
    pages = "251101",
    year = "2010"
}

@article{Xu:2021bfn,
    author = "Xu, Fei and Ezquiaga, Jose Maria and Holz, Daniel E.",
    title = "{Please Repeat: Strong Lensing of Gravitational Waves as a Probe of Compact Binary and Galaxy Populations}",
    eprint = "2105.14390",
    archivePrefix = "arXiv",
    primaryClass = "astro-ph.CO",
    doi = "10.3847/1538-4357/ac58f8",
    journal = "Astrophys. J.",
    volume = "929",
    number = "1",
    pages = "9",
    year = "2022"
}

@article{Ezquiaga:2020spg,
    author = "Ezquiaga, Jose Mar{\'\i}a and Hu, Wayne and Lagos, Macarena",
    title = "{Apparent Superluminality of Lensed Gravitational Waves}",
    eprint = "2005.10702",
    archivePrefix = "arXiv",
    primaryClass = "astro-ph.CO",
    doi = "10.1103/PhysRevD.102.023531",
    journal = "Phys. Rev. D",
    volume = "102",
    number = "2",
    pages = "023531",
    year = "2020"
}

@article{Mukherjee:2019wcg,
    author = "Mukherjee, Suvodip and Wandelt, Benjamin D. and Silk, Joseph",
    title = "{Probing the theory of gravity with gravitational lensing of gravitational waves and galaxy surveys}",
    eprint = "1908.08951",
    archivePrefix = "arXiv",
    primaryClass = "astro-ph.CO",
    doi = "10.1093/mnras/staa827",
    journal = "Mon. Not. Roy. Astron. Soc.",
    volume = "494",
    number = "2",
    pages = "1956--1970",
    year = "2020"
}

@article{Yan:2025,
    author = {Yan, Xingchi and Diebold, Gerald J},
    title = {Determination of Fresnel Integrals for X-ray Phase Contrast Imaging with the Fast Fourier Transform},
    journal = {Microscopy and Microanalysis},
    volume = {31},
    number = {Supplement_1},
    pages = {2311--2313},
    year = {2025},
    month = {07},
    issn = {1431-9276},
    doi = {10.1093/mam/ozaf048.1136},
    url = {https://doi.org/10.1093/mam/ozaf048.1136},
}

@article{Li2020FresnelZonePlate,
  author       = {Ying Li and Ombeline de La Rochefoucauld and Philippe Zeitoun},
  title        = {Simulation of Fresnel Zone Plate Imaging Performance with Number of Zones},
  journal      = {Sensors},
  year         = {2020},
  volume       = {20},
  number       = {22},
  pages        = {6649},
  doi          = {10.3390/s20226649},
  pmid         = {33233576},
  pmcid        = {PMC7699809},
  url          = {https://pubmed.ncbi.nlm.nih.gov/33233576/}
}

@article{Talman1978,
    author  = {Talman, James D.},
    title   = {Numerical Fourier and Bessel transforms in logarithmic variables},
    journal = {Journal of Computational Physics},
    year    = {1978},
    volume  = {29},
    pages   = {35--48},
    doi = {10.1016/0021-9991(78)90107-9},
}

\end{document}